\documentclass{elsart}
\usepackage[dvips]{graphicx,psfrag}
\usepackage{amsfonts}
\usepackage{url}

\begin{document}

\begin{frontmatter}

\title{Estimating short-time period to break different types of chaotic modulation based secure communications\thanksref{hook}}

\author[Spain]{Gonzalo \'{A}lvarez\corauthref{gong}} and
\author[China]{Shujun Li}
\address[Spain]{Instituto de F\'{\i}sica Aplicada, Consejo Superior de
Investigaciones Cient\'{\i}ficas, Serrano 144---28006 Madrid,
Spain}
\address[China]{Department of Electronic Engineering, City University
of Hong Kong, Kowloon Toon, Hong Kong, China}
\corauth[gong]{Corresponding author: Email: gonzalo@iec.csic.es}
\thanks[hook]{Both authors contributed equally to this work.}

\begin{abstract}
In recent years, chaotic attractors have been extensively used in
the design of secure communication systems. One of the preferred
ways of transmitting the information signal is binary chaotic
modulation, in which a binary message modulates a parameter of the
chaotic generator. This paper presents a method of attack based on
estimating the short-time period of the ciphertext generated from
the modulated chaotic attractor. By calculating and then filtering
the short-time period of the transmitted signal it is possible to
obtain the binary information signal with great accuracy without
any knowledge of the parameters of the underlying chaotic system.
This method is successfully applied to various secure
communication systems proposed in the literature based on
different chaotic attractors.
\end{abstract}


\end{frontmatter}

\mathindent 0cm \sloppy

\section{Introduction}

During the last decade, there have been many proposals to apply
non-linear dynamical systems to cryptography and secure
communications under the assumption that chaotic orbits resemble
random-number generators and might mask information signals
\cite{alvarez99,LiThesis2003,yang04}. There exist two main
approaches to chaotic ciphers design: analog and digital. The
first one is based on the concept of chaotic synchronization,
first shown by Pecora and Carrol \cite{pecora90}. In these
systems, the information can be transmitted by the chaotic signal
in a number of ways, including, but not limited to, chaotic
masking \cite{kocarev92,wu93,morgul99,cuomo93a,shahruz02}, in
which the analog message signal $i(t)$ is added to the output of
the chaotic generator $x(t)$ in the transmitter; chaotic switching
or chaos shift keying (CSK) \cite{parlitz92,dedieu93}, in which a
binary message signal is used to choose between two statistically
similar chaotic attractors; chaotic modulation (CM)
\cite{cuomo93b,MHD95,YangCM96,corron97,puebla00,feki03,chen03}, in
which a message, most frequently in binary form, modulates a
parameter of the chaotic generator; or inverse system approach
(ISA), in which the receiver system runs in an exactly inverse way
of the transmitter system to exactly recover the message
\cite{ParlitzAutoSyn96,FeldmannISA96}. Regardless of the method
used to transmit the message signal, the receiver has to
synchronize with the transmitter's chaotic generator to regenerate
the chaotic signal $x(t)$ and thus recover the message $i(t)$.

Chaotic modulation has been repeatedly used as an information
concealing method throughout the years until very recently
\cite{cuomo93b,MHD95,corron97,puebla00,feki03,chen03}. In the
literature, when a binary digital signal is to be encrypted by
this means, one of the state variables of the modulated chaotic
system is commonly taken as the transmitted encrypted message. The
key of the cryptosystem is composed by the unknown internal
parameters of the chaotic system. Thus, the only information
available to the attacker is the instantaneous value of the
transmitted state variable. In \cite{ALLC2004} it is pointed out
that most widely-used chaotic attractors in secure communication
systems exhibit an inherent frequency dependent on the system
parameters. As a result, it is reasonable to assume that the
period of chaotic signals generated using different sets of
parameters must be different. This letter shows that this
assumption proves to be true, even for small variations of the
parameters, and for different types of synchronization and
parameter modulation. A method based on short-time period
estimation is described to detect these slight variations in
period to be able to discern between different attractors and thus
between different values of the binary information signal. The
method works for different chaotic attractors, different
synchronization, and different modulation techniques.

The rest of this letter is organized as follows. In
Sec.~\ref{sec:stp}, the method used to compute the short-time
period is explained. In Sec.~\ref{sec:examples}, some examples of
how the method works are given. The examples use different
modulation techniques to encrypt the message signal. In
Sec.~\ref{sec:comparison}, our method is compared against other
cryptanalytic methods frequently found in the literature. Finally,
Sec.~\ref{sec:conclusion} concludes the letter.

\section{Measuring the short-time period}
\label{sec:stp}

In this section the method followed to calculate the short-time
period of a chaotic scalar signal is explained. It is assumed the
use of 3-D chaotic attractors, given as an autonomous continuous
dynamical system $\dot{\bf x}=\bf f(\bf x)$. Two trajectories
${\bf x}(t)$ and ${\bf x'}(t)$ are said to completely synchronize
if:
\begin{equation}\label{eq:synchronization}
\mathop {\lim }\limits_{t \to \infty } \left| {{\bf x}(t) - {\bf
x'}(t)} \right|= 0.
\end{equation}
For robust synchronization to be maintained, it is required that
all conditional Lyapunov exponents (CLE) of the response subsystem
are negative \cite{pecora90}.

As is well known, chaotic signals present some properties as
sensitive dependence on parameters and initial conditions,
ergodicity, mixing, and dense periodic points. These properties
make them similar to pseudorandom noise. As a result, this
apparent randomness has motivated their use in secure
communication applications. The most widely-used chaotic signal
generators in this context are based on the double-scroll Lorenz
and Chua attractors, and on the single-scroll R\"{o}ssler
attractor. As studied in \cite{ALLC2004}, these chaotic attractors
exhibit an inherent frequency uniquely determined by their system
parameters. When present, this frequency can be measured over
long-time periods. However, in this work we are interested in
knowing the fast fluctuations in the frequency in short term in an
effort to estimate the attractor's instantaneous frequency. We try
to measure the short-time period as a function of time to unmask
the binary modulating signal. If the signal is periodic or
nearly-periodic, calculating the short-time zero-crossing rate
(STZCR) or a short Discrete Fourier Transform (DFT) would be
enough, but chaotic signals are essentially aperiodic.
Nevertheless, along the trajectory followed by an initial point in
these attractors there are regions where the movement is very
close to periodic, thus allowing for a very accurate estimation of
the short-time period. The peculiarities of different chaotic
signals require the customization of the method for the three
different types of attractors under consideration. Once a
sufficiently stable region is found as described in the next
section, then it is possible to compute the short-time period
following the procedure described below.

In parameter modulation based chaotic secure communication
systems, one parameter of the attractor is changed according to
the binary value of the information signal $i(t)$ regardless of
the synchronization method. Usually one state variable of the
attractor, $x_i(t)$, is used to convey the concealed information
signal. This state variable is the only information available to
the attacker. Let us note that the orbit followed by an initial
condition is generally non-uniform. However, if a 2-D projection
of the chaotic attractor is used, it will be observed that there
is always a region in which the average rotation angular speed is
almost constant, \emph{i.e.}, the elapsed time for each visit to
this region is almost constant. The part of the signal
corresponding to this nearly-periodic region is used to make the
measure as accurate as possible. Thus, instead of measuring the
whole period, which may lead to inaccurate results, only a
fraction of the period is measured, corresponding to the elapsed
time within the nearly-periodic region in each rotation. Let us
note that this method tries to spot variations in the short-time
period and is not concerned with measuring its exact value. Next,
a new time signal $p(t)$ is created, by assigning this measured
value to $p(t)$ for the duration of the whole rotation period of
$x_i(t)$. Once $p(t)$ has been created, its DC component is
removed by subtracting its mean value. The new signal is $p^*(t)$.
Last, an appropriate moving averaging filter with a Hanning window
to smooth up the result is used on $p^*(t)$. As will be seen, the
resulting filtered signal, $fp^*(t)$, suffices to detect the
plaintext, although a Schmitt-Trigger with adequate switch-on and
switch-off levels might be used to obtain the final recovered
signal, $i^*(t)$.

In the next section, several examples are given where the above
process is further explained and successfully applied to the
cryptanalysis of different types of chaotic modulation based
secure communication systems.

\section{Examples}
\label{sec:examples}

In this section the performance of the short-time period
estimation method is analyzed when applied to different secure
communication systems proposed in past and recent literature,
including the classical parameter modulation method, a phase
synchronization method, and an adaptive observer-based chaos
synchronization method. We believe such a cryptanalysis method can
be further generalized to break other secure communication
systems.

\subsection{Classical parameter modulation method}

The first implementation of parameter modulation \cite{cuomo93b}
uses the well-known double-scroll Lorenz attractor \cite{lorenz63}
as the chaotic signal generator. The transmitter end is
represented by:
\begin{eqnarray}
&&\dot{x}_1=\sigma(x_2-x_1),\nonumber\\
&&\dot{x}_2=rx_1 - x_2 - x_1x_3,\\
&&\dot{x}_3=x_1x_2 - b(i(t))x_3,\nonumber
\end{eqnarray}
where $\sigma$, $r$, and $b$ are the internal system parameters.
Furthermore, $i(t)$ is a binary information signal controlling the
parameter $b$ to be one of two different values $b_0$ and $b_1$.
At the receiver end, an identical system is used by tuning the
parameter $b$ to be $b_0$ (or $b_1$). When the receiver subsystem
synchronizes with the transmitted signal in the sense described by
Eq.~(\ref{eq:synchronization}), it is known that $b_0$ (or $b_1$)
was used at the transmitter end; when it does not synchronize, the
other value is assumed. In such a way, the binary message $i(t)$
is decrypted from $b=b_0$ or $b_1$ at each given time $t$.

\begin{figure}
  \centering
  \psfrag{it}{$i(t)$}
  \psfrag{x1}{$x_1(t)$}
  \psfrag{pt}{$p(t)$}
  \psfrag{p1t}{$p^*(t)$}
  \psfrag{fpt}{$fp^*(t)$}
  \psfrag{i1t}{$i^*(t)$}
  \psfrag{time(sec)}{time (sec)}
  \includegraphics{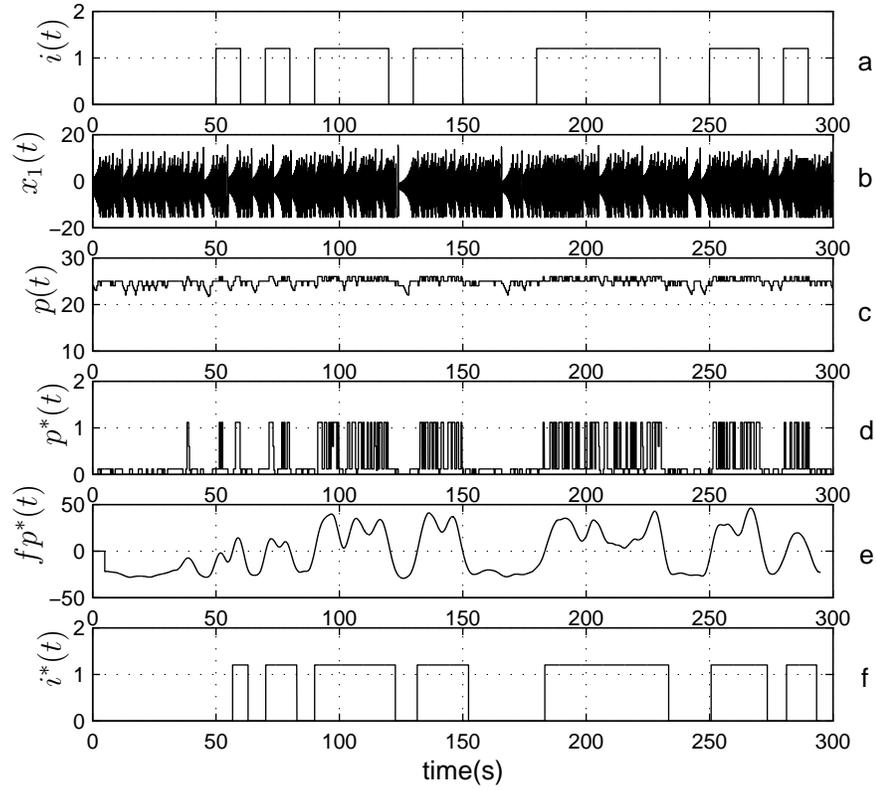}
  \caption{Breaking classical parameter modulation using Lorenz attractor:
  a) original binary information signal, $i(t)$;
  b) the transmitted state variable signal or ciphertext, $x_1(t)$;
  c) the short-time period signal, $p(t)$;
  d) the positive value after removing DC component, $p^*(t)$;
  e) the low-pass filtered signal, $fp^*(t)$, revealing the modulation signal;
  f) recovered message signal, $i^*(t)$, after adequate detection. }
  \label{fig:LorenzBreak}
\end{figure}

When the Lorenz attractor is used, usually either $x_1(t)$ or
$x_2(t)$ is transmitted as ciphertext. In \cite{cuomo93b},
$x_1(t)$ is used (see Fig.~\ref{fig:LorenzBreak}.b),
$\sigma=16.0$, $r=45.6$, and the modulated parameter is $b$,
taking values $b=4$ or 4.4 for the binary signal equal to $i(t)=0$
or $i(t)=1$, respectively. The message signal $i(t)$ is plotted in
Fig.~\ref{fig:LorenzBreak}.a.

\begin{figure}[h]
  \center
  \psfrag{Quasi-periodic}{}
  \psfrag{x1}{$x_1$}\psfrag{x12}{$|x_1|$}\psfrag{x3}{$x_3$}\psfrag{K}{}
  \includegraphics{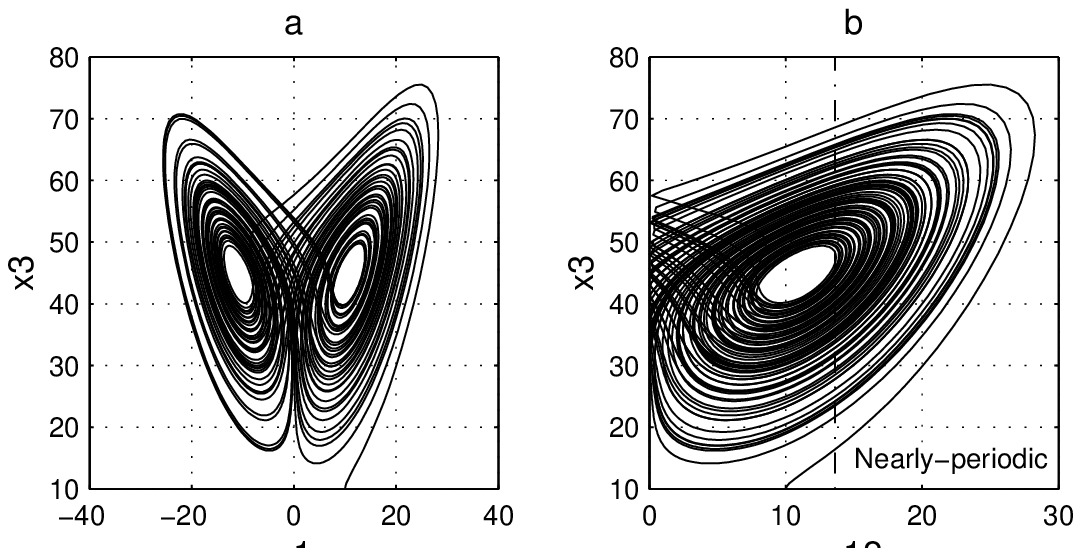}
  \caption{The Lorenz attractor: a) $x_1-x_3$ projection;
  b) $|x_1|-x_3$ projection.}
  \label{fig:LorenzAttractor}
\end{figure}

The Lorenz chaotic signal must be first correctly conditioned
before computing its short-time period. As can be observed in
Fig.~\ref{fig:LorenzAttractor}.a, the orbit followed by an
arbitrary initial point spirals around the two scrolls, jumping
from one scroll to the other in a chaotic manner. The work with
this attractor is simplified if its absolute value is taken:
$y_i(t)=|x_i(t)|$, $i=\{1,2\}$. This operation folds the attractor
back on itself due to its symmetry with respect to $x_1=0$ and
$x_2=0$, in such a way that the trajectories spiral around one
merged scroll in the same rotation direction, as observed in
Fig.~\ref{fig:LorenzAttractor}.b. As discussed in the previous
section, to obtain the maximum accuracy in the estimations, the
rotation duration is measured on the region where the average
rotation angular speed is almost constant. This region, plotted in
Fig.~\ref{fig:LorenzAttractor}.b, can be easily computed as the
region to the right of the middle value of the maximum value of
$y_i(t)$. In the example, the period is computed as the elapsed
time during which $\max(y_1)/2<y_1(t)<\max(y_1)$ holds.

The rest of the process of measuring the short-time period value
is quite similar to the one outlined in the previous section, but
proceeding with $y_i(t)$ instead. Following this process, plotted
in Figs.~\ref{fig:LorenzBreak}.c-f, the original message signal is
recovered with great accuracy. A Schmitt-Trigger with switch-on
level of 0 and switch-off level of $-20$ was used. The recovered
signal $i^*(t)$ is slightly delayed with respect to the original
$i(t)$ due to the delay introduced by the filter and can be easily
removed if desired.

It must be added that this method of parameter modulation has been
known to be insecure many years before
\cite{yang95,yang98a,yang98b,yang98c}.

\subsection{Phase synchronization method}

Most secure chaotic communication systems are based on complete
synchronization in the sense of Eq.~(\ref{eq:synchronization}),
whereas new cryptosystems have been proposed based on phase
synchronization \cite{chen03}. This scheme hides binary messages
in the instantaneous phase of the drive subsystem used as the
transmitting signal to drive the response subsystem. At the
receiver, the phase difference is detected and its strong
fluctuation above or below zero allows the plaintext recovering at
certain coupling strength. The secure communication process is
illustrated in \cite{chen03} by means of an example based on
coupled R\"ossler chaotic oscillators. In the example, the drive
subsystem is formed by two weakly-coupled oscillators. The
plaintext is used to modulate the same parameter in both
oscillators 1 and 2. The equations of the drive subsystem are:
\begin{eqnarray}
&&\dot{x}_{1,2}=-(\omega+\Delta\omega)y_{1,2}-z_{1,2}+
\varepsilon(x_{2,1}-x_{1,2}),\nonumber\\
&&\dot{y}_{1,2}=(\omega+\Delta\omega)x_{1,2}+\alpha y_{1,2},\\
&&\dot{z}_{1,2}=0.2+z_{1,2}(x_{1,2}-10).\nonumber
\end{eqnarray}
The response subsystem is governed by:
\begin{eqnarray}
&&\dot{x}_{3}=-\omega\,'y_{3}-z_{3}+
\eta((x^2_3+y_3^2)^{1/2}cos\phi_m-x_3),\nonumber\\
&&\dot{y}_{3}=\omega\,'x_{3}+\alpha\,' y_{3},\\
&&\dot{z}_{3}=0.2+z_{3}(x_{3}-10).\nonumber
\end{eqnarray}
In the example, the parameter values are: $\omega=\omega\,'=1,$
$\varepsilon=5\times10^{-3}$, $\eta=5.3$, and
$\alpha=\alpha\,'=0.15\,.$ The parameter $\omega$ corresponds to
the natural frequency of the R\"{o}ssler oscillator drive
subsystems 1 and 2. The parameter $\omega\,'$ corresponds to the
natural frequency of the R\"{o}ssler oscillator driven subsystem
3, $\varepsilon$ corresponds to the weak coupling factor between
the oscillators 1 and 2, and $\eta$ corresponds to the strong
coupling factor between the 2 driven oscillators and the response
oscillator 3. The parameter mismatch $\Delta\omega$ is modulated
by the plaintext, being $\Delta\omega=0.01$ if the bit to be
transmitted is ``1'' and $\Delta\omega=-0.01$ if the bit to be
transmitted is ``0''.

The ciphertext consists of the phase of the mean field of the
drive oscillators:
\begin{equation}\label{eq:phase}
  \phi_m=\arctan\frac{x_1+x_2}{y_1+y_2},
\end{equation}
where $\arctan$ is the arctangent function of the argument, from
$-\pi$ to $\pi$.

\begin{figure}
  \centering
  \psfrag{x1}{$x_1$}\psfrag{x2}{$y_1$}\psfrag{x3}{$z_1$}
  \includegraphics{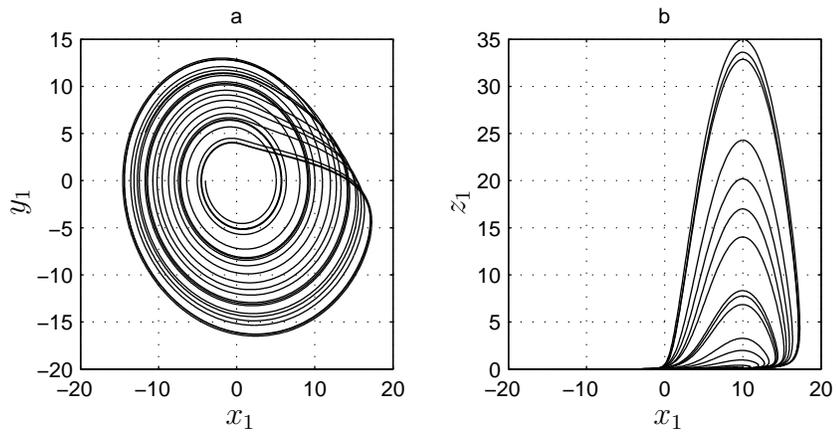}
  \caption{The well-known R\"{o}ssler attractor:
  a) $x_1-y_1$ projection;
  b) $x_1-z_1$ projection.}
  \label{fig:RosslerAttractor}
\end{figure}

\begin{figure}
  \centering
  \psfrag{it}{$i(t)$}
  \psfrag{phi}{$\phi_m(t)$}
  \psfrag{pt}{$p(t)$}
  \psfrag{p1t}{$p^*(t)$}
  \psfrag{i1t}{$i^*(t)$}
  \psfrag{time}{time(sec)}
  \includegraphics{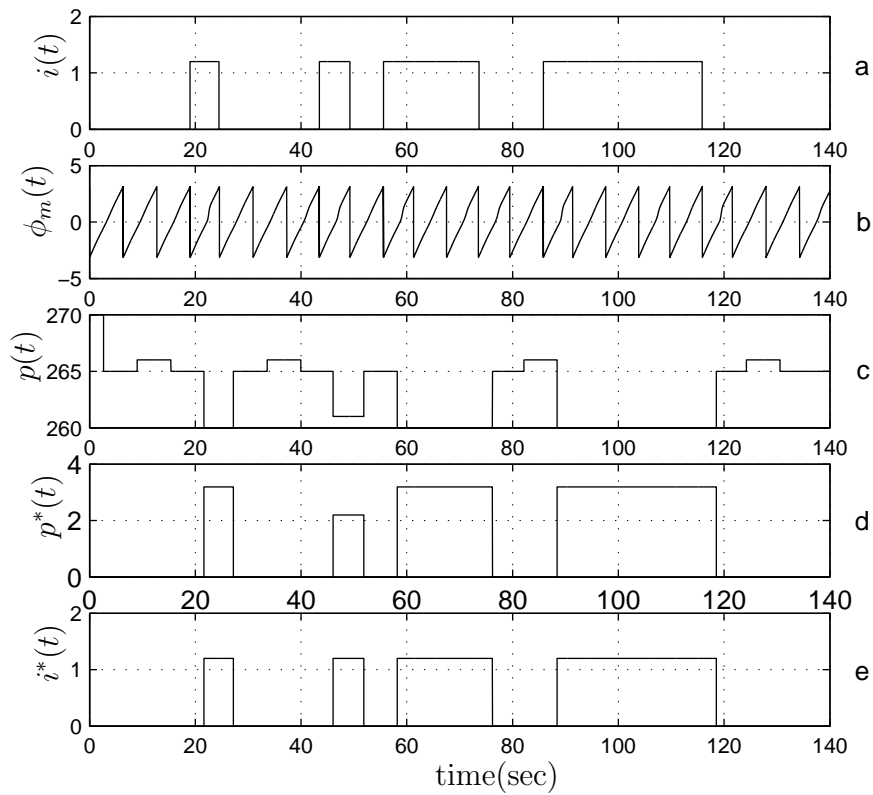}
  \caption{Breaking phase synchronization using R\"{o}ssler attractor:
  a) original binary information signal, $i(t)$;
  b) the transmitted phase signal or ciphertext, $\phi_m(t)$;
  c) the short-time period signal, $p(t)$;
  d) the positive value after removing DC component, $p^*(t)$;
  e) recovered message signal, $i^*(t)$, after adequate detection. }
  \label{fig:RosslerBreak}
\end{figure}

The signal available to the attacker is $\phi_m(t)$, the
instantaneous phase. In the following, without loss of generality,
only $\phi_1(t)=\arctan(x_1/y_1)$ is considered to qualitatively
illustrate the behavior of the R\"ossler attractor. In
Fig.~\ref{fig:RosslerAttractor}.a it is observed that for
$x_1(t)<0$, the rotation angular speed is approximately constant,
\emph{i.e.}, the phase increases almost linearly. However,
depending on the system parameters chosen, the phase can change
abruptly in the first quadrant when $0<\phi_1(t)<\pi/2$. Thus,
this is the part of the signal to be avoided to compute the
short-time period. Although this cryptosystem was already broken
by an economic brute-force attack in \cite{Alvarez04a},
Fig.~\ref{fig:RosslerBreak} shows the results obtained after
applying a more elegant and straightforward avenue of attack using
the cryptanalysis described in this letter. In the example
analyzed, the signal conditioning is limited to considering
$\pi/2<|\phi_m(t)|<\pi$ to compute $p(t)$. In this case, it is not
necessary to filter $p^*(t)$ because each bit of the plain-signal
corresponds exactly to one short-time period of $p^*(t)$. Thus, by
simply rescaling $p^*(t)$ a perfect estimation of $i(t)$ is
obtained. Again, the time delay can be removed if desired.

\subsection{Adaptive observer-based chaos synchronization}

In \cite{feki03}, the author proposes a symmetric secure
communication system based on parameter modulation of a chaotic
oscillator acting as a transmitter. The receiver is a chaotic
system synchronized by means of an adaptive observer. Two sample
implementations are given: one with the Lorenz attractor and
another with Chua attractor. In this letter the latter will be
broken, to illustrate how our method works with a different
double-scroll attractor. It works equally well for Lorenz, though.

Chua's circuit dynamics can be described by the following
equations:
\begin{eqnarray}
&&\dot{x}_1=\alpha(-x_1+x_2)-f_1(x_1),\nonumber\\
&&\dot{x}_2=x_1-x_2+x_3,\\
&&\dot{x}_3=-\beta x_2.\nonumber
\end{eqnarray}
where $f_1(x)=bx+0.5(a-b)(|x+1|-|x-1|)$. In the example the system
is implemented with the following parameter values,
$(\alpha,\beta,a,b)=(10,18,-4/3,-3/4)$. The signal used for
synchronization of the receiver is $x_1$. The encryption process
is defined by modulating the parameter $\beta$ with the binary
encoded plaintext, so that it is $\beta+1.25$ if the plaintext bit
is "1" and $\beta-1.25$ if the plaintext bit is "0". The duration
of the plaintext bits must be much larger than the convergence
time of the adaption law. The uncertain system can be rewritten in
a compact form as:
\begin{equation}\label{eq:matrix1}
\left[ {\begin{array}{*{20}c}
   {\dot x_1 }  \\
   {\dot x_2 }  \\
   {\dot x_3 }  \\
\end{array}} \right] = \left[ {\begin{array}{*{20}c}
   -10 & 10 & 0  \\
   1 & -1 & 1  \\
   0 & \beta & 0  \\
\end{array}} \right]\left[ {\begin{array}{*{20}c}
   x_1  \\
   x_2  \\
   x_3  \\
\end{array}} \right] + \left( {\begin{array}{*{20}c}
   f_1(x_1)  \\
   0 \\
   0 \\
\end{array}} \right) + \left[ {\begin{array}{*{20}c}
   0  \\
   0  \\
   1  \\
\end{array}} \right]x_2\theta,
\end{equation}

\begin{equation}\label{eq:y}
y=C\cdot x=x_3,
\end{equation}
\begin{equation}\label{eq:C}
C=[0~0~1],
\end{equation}
\begin{equation}\label{eq:teta}
\theta=\Delta\beta=\pm 1.25.
\end{equation}
The transmitted ciphertext is the signal $x_3(t)$.

Again, we are dealing with a double-scroll attractor. In contrast
to the Lorenz attractor, the trajectory followed by an arbitrary
initial point in the Chua attractor is uniform enough to allow the
direct estimation of the (whole) short-time period from the
transmitted signal $x_3(t)$.

\begin{figure}
  \centering
  \psfrag{it}{$i(t)$}
  \psfrag{x3t}{$x_3(t)$}
  \psfrag{pt}{$p(t)$}
  \psfrag{p1t}{$p^*(t)$}
  \psfrag{fpt}{$fp^*(t)$}
  \psfrag{i1t}{$i^*(t)$}
  \psfrag{time}{time(sec)}
  \includegraphics{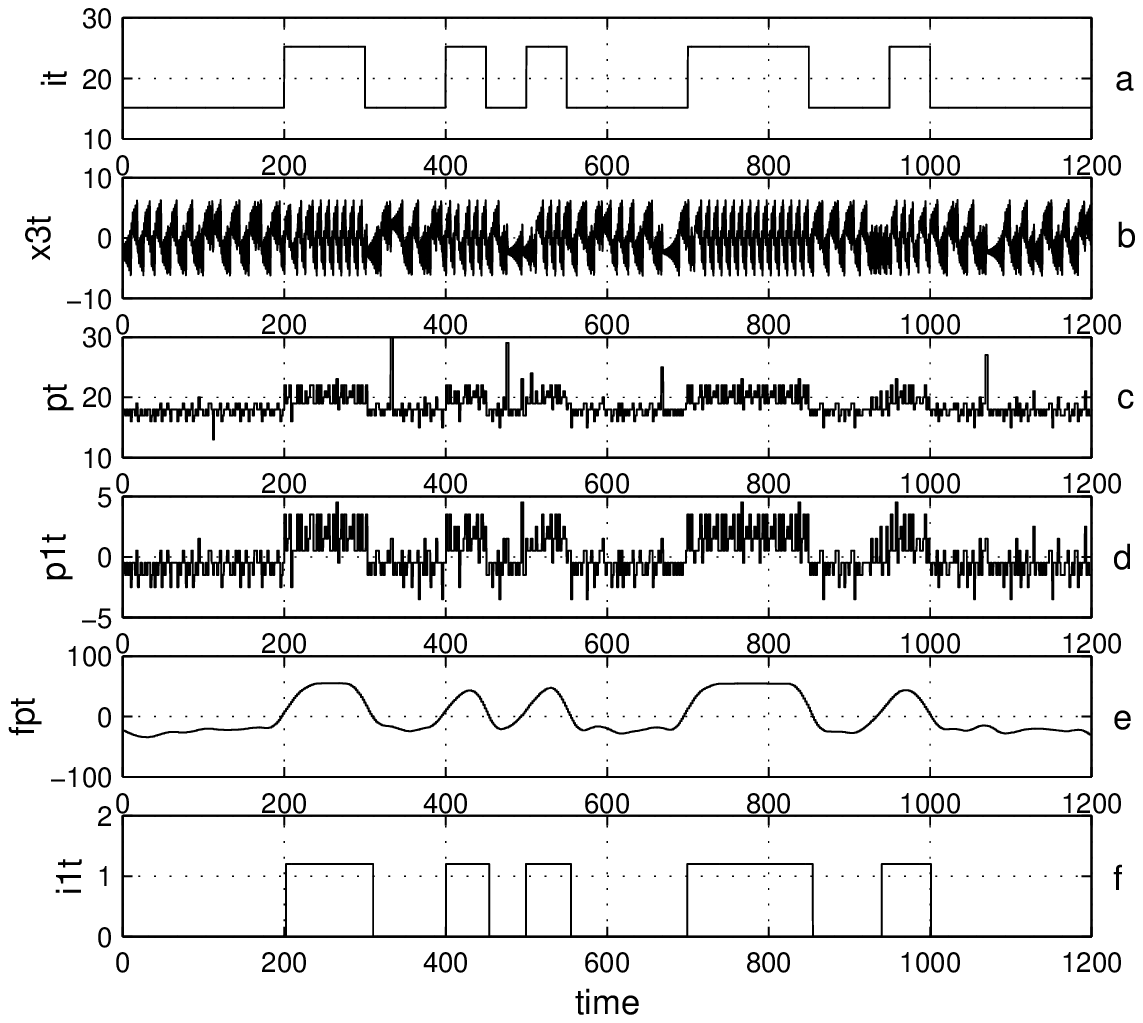}
  \caption{Breaking adaptive observer-based chaos synchronization using Chua attractor:
  a) original binary information signal, $i(t)$;
  b) the transmitted state variable signal or ciphertext, $x_3(t)$;
  c) the short-time period signal, $p(t)$;
  d) the clipped signal, $p^*(t)$, after removing singular peaks and DC component;
  e) the low-pass filtered signal, $fp^*(t)$, revealing the modulation signal;
  f) recovered message signal, $i^*(t)$, after adequate detection.}
  \label{fig:ChuaBreak}
\end{figure}

After applying this method to $x_3(t)$, as shown in
Figs.~\ref{fig:ChuaBreak}.c-f, the original message signal is
recovered with great accuracy. A Schmitt-Trigger with switch-on
and switch-off levels of 10 and 0 respectively was used. The
recovered signal $i^*(t)$ is slightly delayed with respect to the
original $i(t)$ due to the delay introduced by the filter and can
be easily removed if desired.

\section{Comparison with other attack methods}
\label{sec:comparison}

Throughout the years, different methods have been proposed to
attack chaos-based secure communication systems. In this section,
the performance of the short-time period estimation method is
compared against the most relevant.

The return-map method was initially devised by \cite{perez95} and
further developed by \cite{yang98a}. Given one of the variables in
the chaotic system, one or more proper return maps can be
constructed allowing for a partial reconstruction of the dynamics.
By analyzing the evolution of the signal on the attracting sets of
those maps, the message can be extracted under certain conditions.
These attacks can be performed without the knowledge of the
precise structure of the chaotic system in use. This method not
only decrypts ciphertexts encrypted using chaotic modulation, but
also using chaotic masking. However, it does not work for phase
synchronization cryptosystems. There are some improved
cryptosystems \cite{BuWang04} which avoid the return map attack by
modulating the transmitted signal with an appropriately chosen
scalar signal. Our method was checked against this improved
method. The results show that it is still able to directly recover
the correct signal, and also can be used to identify and remove
the modulating signal, thus rendering the return-map attack again
possible.

In \cite{yang95}, the short-time zero-crossing rate (STZCR) of the
differential of the transmitted signal is used to recover the
information digital signal. This method presents the limitation of
only working on single-scroll Chua's circuits proposed in
\cite{parlitz92}, while this letter generalizes this method so
that can be used on different chaotic attractors, including the
three most frequently-used ones, \emph{i.e.}, (double-scroll)
Lorenz, (single-scroll) R\"{o}ssler and (double-scroll) Chua
attractors. In this letter different conditioning methods are
discussed to show the great flexibility of our method. In doing
so, we have partially revealed the theory hiding behind the
nearly-stable short-time period of many 3-D chaotic attractors.

When two different attractors (or for the same event, the same
attractor with two different parameter sets) are switched to
encode a binary message, a spectrogram might reveal the evolution
of the energy distribution in spectral-time space from the
transmitted signal. If the two chaotic attractors have some
detectable difference in their spectrums, then the spectrogram can
be used to detect this difference and thus unmask the scrambled
binary information \cite{yang98b}. This method can be used in
chaotic masking too, but does not work for phase synchronization.

In the same way that changing the parameter in an attractor
affects its frequency, it is reasonable to assume that also small
changes in its amplitude will take place when shifting from one
set of parameters to the other. This approach was used in
\cite{Alvarez04b}, squaring the ciphertext signal and low-pass
filtering it, so that the enveloping waveform, \emph{i.e.}, the
binary modulating signal, was finally extracted. This method
performs well when the difference in amplitude of the two bits in
the modulating square waveform is big enough to be observed after
the filtering. Obviously, it does not work for phase
synchronization where the amplitude does not change.

The generalized synchronization attack, first introduced by
\cite{yang98c}, assumes that the attacker knows the type of
attractor used for the transmission and reception, but ignores the
precise value of the parameters, which usually are considered to
be the secret key of the cryptosystem. Using the concept of
generalized synchronization (GS) defined in \cite{Rulkov95}, the
attacker's receiver uses a set of parameters which is completely
different to the secret key and thus will never achieve
synchronization. Nevertheless, by measuring the synchronization
error over time, it is possible to detect the switching between
the two attractors in the transmitter as a variation in the square
error. This one is a very powerful technique when complete
synchronization is used. It doesn't work for some other types of
synchronization though.

\section{Conclusion}
\label{sec:conclusion}

A new cryptanalytic method to break parameter modulation based
chaotic secure communication systems is presented. The method
computes the short-time period of the ciphertext signal to detect
slight variations in its frequency. For the method to work in a
wide variety of modulation techniques and for different chaotic
attractors, first the transmitted signal must be conditioned
according to the structure of the underlying chaotic attractor
used for the modulation. The letter describes the different
conditioning processes required for different attractors and
explains how to calculate the short-time period variation as a
function of time of the conditioned signal. The signal processing
required to eventually recover the original plaintext is
explained. Finally, this method is compared to some other
cryptanalytic techniques used in literature. It is shown that it
is the first method apart from brute force which recovers the
signal when phase synchronization is used. Some important facts
about the nearly-stable short-time period of many 3-D chaotic
attractors are also revealed by this work.

\ack{This work is supported by Ministerio de Ciencia y
Tecnolog\'{\i}a of Spain, research grants TIC2001-0586 and
SEG2004-02418, and by the Applied R\&D Center, City University of
Hong Kong, Hong Kong SAR, China, under Grants no. 9410011 and no.
9620004.}

\clearpage

\section*{Figures and captions}

\clearpage

\clearpage

\clearpage

\clearpage

\begin{thebibliography}{10}

\bibitem{alvarez99}
G.~\'{A}lvarez, F.~Montoya, M.~Romera, and G.~Pastor.
\newblock Chaotic cryptosystems.
\newblock In Larry~D. Sanson, editor, {\em 33rd Annual 1999 International
  Carnahan Conference on Security Technology}, pages 332--338. IEEE, 1999.

\bibitem{LiThesis2003}
Shujun Li.
\newblock {\em Analyses and New Designs of Digital Chaotic Ciphers}.
\newblock PhD thesis, School of Electronics and Information Engineering, Xi'an
  Jiaotong University, Xi'an, China, June 2003.
\newblock Available online at \url{http://www.hooklee.com/pub.html}.

\bibitem{yang04}
T.~Yang.
\newblock A survey of chaotic secure communication systems.
\newblock {\em Int. J. Comp. Cognition}, 2(2):81--130, 2004.

\bibitem{pecora90}
L.~M. Pecora and T.~L. Carroll.
\newblock Synchronization in chaotic systems.
\newblock {\em Phys. Rev. Lett.}, 64(8):821--824, 1990.

\bibitem{kocarev92}
L.~Kocarev, K.~S. Halle, K.~Eckert, L.~O. Chua, and U.~Parlitz.
\newblock Experimental demonstration of secure communications via chaotic
  synchronization.
\newblock {\em Int. J. Bifur. \textup{\&} Chaos}, 2(3):709--713, 1992.

\bibitem{wu93}
C.~W. Wu and L.~O. Chua.
\newblock A simple way to synchronize chaotic systems with applications to
  secure communications systems.
\newblock {\em Int. J. Bifur. \textup{\&} Chaos}, 3(6):1619--1627, 1993.

\bibitem{morgul99}
Omer Morgul and Moez Feki.
\newblock A chaotic masking scheme by using synchronized chaotic systems.
\newblock {\em Phys. Lett. A}, 251(3):169--176, 1999.

\bibitem{cuomo93a}
K.~M. Cuomo, A.~V. Openheim, and S.~H. Strogatz.
\newblock Synchronization of {Lorenz}-based chaotic circuits with applications
  to communications.
\newblock 40(10):626--633, 1993.

\bibitem{shahruz02}
S.~M. Shahruz, A.~K. Pradeep, and R.~Gurumoorthy.
\newblock Design of a novel cryptosystem based on chaotic oscillators and
  feedback inversion.
\newblock {\em J. Sound and Vibration}, 250(4):762--771, 2002.

\bibitem{parlitz92}
U.~Parlitz, L.~O. Chua, L.~Kocarev, K.~S. Halle, and A.~Shang.
\newblock Transmission of digital signals by chaotic synchronization.
\newblock {\em Int. J. Bifur. \textup{\&} Chaos}, 2(4):973--977, 1992.

\bibitem{dedieu93}
H.~Dedieu, M.~P. Kennedy, and M.~Hasler.
\newblock Chaos shift keying: Modulation and demodulation of a chaotic carrier
  using self-synchronizing.
\newblock 40(10):634--641, 1993.

\bibitem{cuomo93b}
K.~M. Cuomo and A.~V. Openheim.
\newblock Circuit implementation of synchronized chaos with applications to
  communications.
\newblock {\em Phys. Rev. Lett.}, 71(1):65--68, 1993.

\bibitem{MHD95}
M.~Martinenssen, B.~H\"{u}binger, and R.~Doerner.
\newblock Chaotic cryptology.
\newblock {\em Ann. Physik}, 4(1):35--42, 1995.

\bibitem{YangCM96}
T.~Yang and L.~O. Chua.
\newblock Secure communication via chaotic parameter modulation.
\newblock 43(9):817--819, 1996.

\bibitem{corron97}
N.~J. Corron and D.~W. Hahs.
\newblock A new approach to communications using chaotic signals.
\newblock 44(5):373--382, 1997.

\bibitem{puebla00}
H.~Puebla and J.~Alvarez-Ramirez.
\newblock Proportional-integral feedback demodulation for secure
  communications.
\newblock {\em Phys. Lett. A}, 276(5-6):245--256, 2000.

\bibitem{feki03}
M.~Feki.
\newblock An adaptive chaos synchronization scheme applied to secure
  communication.
\newblock {\em Chaos, Solitons \textup{\&} Fractals}, 18(1):141--148, 2003.

\bibitem{chen03}
J.~Y. Chen, K.~W. Wong, L.~M. Cheng, and J.~W. Shuai.
\newblock A secure communication scheme based on the phase synchronization of
  chaotic systems.
\newblock {\em Chaos}, 13(2):508--514, 2003.

\bibitem{ParlitzAutoSyn96}
U.~Parlitz and L.~Kocarev.
\newblock Multichannel communication using autosynchronization.
\newblock {\em Int. J. Bifur. \textup{\&} Chaos}, 6(3):581--588, 1996.

\bibitem{FeldmannISA96}
U.~Feldmann, M.~Hasler, and W.~Schwarz.
\newblock Communication by chaotic signals: The inverse system approach.
\newblock {\em Int. J. Circuit Theory Appl.}, 24(5):551--579, 1996.

\bibitem{ALLC2004}
G.~\'{A}lvarez, S.~Li, J.~L\"{u}, and G.~Chen.
\newblock Inherent frequency and spatial decomposition of the {Lorenz} chaotic
  attractor.
\newblock arXiv: nlin.CD/0406031, 2004.

\bibitem{lorenz63}
E.~N. Lorenz.
\newblock Deterministic nonperiodic flow.
\newblock {\em J. Atmosph. Sc.}, 20(2):130--141, 1963.

\bibitem{yang95}
T.~Yang.
\newblock Recovery of digital signals from chaotic switching.
\newblock {\em Int. J. Circuit Theory Appl.}, 23(6):611--615, 1995.

\bibitem{yang98a}
T.~Yang, L.~B. Yang, and C.~M. Yang.
\newblock Cryptanalyzing chaotic secure communications using return maps.
\newblock {\em Phys. Lett. A}, 245(6):495--510, 1998.

\bibitem{yang98b}
T.~Yang, L.~B. Yang, and C.~M. Yang.
\newblock Breaking chaotic secure communications using a spectogram.
\newblock {\em Phys. Lett. A}, 247(1-2):105--111, 1998.

\bibitem{yang98c}
T.~Yang, L.~B. Yang, and C.~M. Yang.
\newblock Breaking chaotic switching using generalized synchronization:
  Examples.
\newblock 45(10):1062--1067, 1998.

\bibitem{Alvarez04a}
G.~\'{A}lvarez, F.~Montoya, M.~Romera, and G.~Pastor.
\newblock Breaking a secure communication scheme based on the phase
  synchronization of chaotic systems.
\newblock {\em Chaos}, 14(2):274--278, 2004.

\bibitem{perez95}
G.~P\'{e}rez and H.~A. Cerdeira.
\newblock Extracting messages masked by chaos.
\newblock {\em Phys. Rev. Lett.}, 74(11):1970--1973, 1995.

\bibitem{BuWang04}
S.~Bu and B.-H. Wang.
\newblock Improving the security of chaotic encryption by using a simple
  modulating method.
\newblock {\em Chaos, Solitons \textup{\&} Fractals}, 19(4):919--924, 2004.

\bibitem{Alvarez04b}
G.~\'{A}lvarez, F.~Montoya, M.~Romera, and G.~Pastor.
\newblock Breaking parameter modulated chaotic secure communication system.
\newblock {\em Chaos, Solitons \textup{\&} Fractals}, 21(4):783--787, 2004.

\bibitem{Rulkov95}
N.~F. Rulkov, M.~M. Sushckik, L.~S. Tsimring, and H.~D.~I.
Abarbanel.
\newblock Generalized syncronization of chaos in directionally coupled chaotic
  systems.
\newblock {\em Phys. Rev. E}, 51(2):980--994, 1995.

\end{thebibliography}
\end{document}